\documentclass[amsmath, amsfonts, showpacs, twocolumn, superscriptaddress, prb]{revtex4}
\usepackage{graphicx}
\usepackage{epsfig}
\usepackage{bm}
\usepackage{dcolumn}
\usepackage{color}
\usepackage{amsmath}
\usepackage{amssymb}
\begin{document}

\title{Singular length dependence of critical current in SNS bridges}
\author{Alex Levchenko}
\affiliation{Department of Physics, University of Minnesota,
Minneapolis, MN 55455, USA}
\author{Alex Kamenev}
\affiliation{Department of Physics, University of Minnesota,
Minneapolis, MN 55455, USA}
\author{Leonid Glazman}
\affiliation{Department of Physics, University of Minnesota,
Minneapolis, MN 55455, USA} \affiliation{W.I. Fine Theoretical
Physics Institute, University of Minnesota, Minneapolis, MN 55455,
USA}
\date{October 25, 2006}
\begin{abstract}
  We examine dependence of the critical Josephson current on the
  length $L$ of the normal bridge N between two bulk superconductors.
  This dependence turns out to be nonanalytic at small $L$. The
  nonanalyticity originates from the contribution of extended
  quasiparticle states with energies well above the superconducting
  gap. This should be contrasted with the more familiar contribution
  to the Josephson current coming from Andreev bound states localized
  in the normal region at energies below the gap.
\end{abstract}
\maketitle

Owing to the rapid progress in
nanofabrication~\cite{Nano1,Nano2,Nano3,Nano4}, there is a renewed
interest in various aspects of the physics of
superconductor/normal-metal/superconductor (SNS)
structures~\cite{Review}. The purpose of this paper is to point out
a subtle contribution to the Josephson current, flowing across the
normal region. Specifically, we address the current carried by the
high energy, $\varepsilon>\Delta$, {\em extended} states of the
system. It leads to a nonanalytic behavior of the critical current
as a function of the length of the normal region, $L$, at small $L$.

It is frequently stated \cite{Beenakker,Heikkila} that as the result
of Andreev reflection~\cite{Andreev} the Josephson current is
carried exclusively by the Andreev bound states with the energies
$\varepsilon < \Delta$, {\em localized} in the normal region.
Indeed, the energies of such bound states are sensitive to the phase
difference, $\phi$, between the superconductors. This leads to the
$\phi$--dependent free energy and thus to the supercurrent. The
validity of this point of view is well established in the two
limiting cases of long \cite{Schon,Wilhelm,Zaikin} junction $L\gg
\xi$, and very short \cite{Beenakker,Heikkila,KO} one, $L/\xi\to 0$,
where $\xi$ is the coherence length of the superconductor. Thus it
came as a surprise for us that in between these two extremes, the
physics appears to be more complicated. We find that the Josephson
current is shared between the localized Andreev states and the
extended above--gap states. The dependence of the critical current
on $L$ at short lengths, $L\ll\xi$, is not analytic and comes from
the contribution of the extended states.

The phase sensitivity of the energy levels originates from the
trajectories, which are reflected at least once from both NS
interfaces. Such trajectories, require propagation time longer than
the diffusion time across the normal region $L^2/D$, where $D$ is
the diffusion constant. As a result, only states with energy smaller
than the Thouless energy, $\varepsilon\lesssim E_{\rm Th}=\hbar
D/L^2$, may exhibit sensitivity to the phase difference. In the long
junction limit, the Thouless energy is small compared to the
superconducting gap, $E_{\rm Th}\ll \Delta$ (indeed, notice that
$\Delta=\hbar D/\xi^2$). Thus all the current--carrying states are
at the energy $\varepsilon\lesssim E_{\rm Th}\ll \Delta$.  Since
there are no states in the superconducting leads at such energies,
the Andreev bound states of the normal region are indeed solely
responsible for the Josephson current.

The situation is more complicated for shorter junctions, such that
$E_{\rm Th}>\Delta$. One may expect that the extended states in the
energy range $\Delta<\varepsilon < E_{\rm Th}$ contribute to the
supercurrent. Yet, in the only case treated analytically so far,
namely in the limit $L/\xi\to 0$ (or, equivalently, $E_{\rm Th}\to
\infty$) the entire current is still given by the Andreev states
\cite{KO,Beenakker} with $0<\varepsilon\leq \Delta$. In this paper,
we show that this result is an artefact of the limit $E_{\rm Th}\to
\infty$ rather than a general principle. Specifically, the spectral
density of the Josephson current appears to be $J(\varepsilon)\sim
\Delta^2/(E_{\rm
  Th}\varepsilon)$ in the energy interval $\Delta\lesssim \varepsilon
\lesssim E_{\rm Th}$. Upon integration over energy, it results in
the contribution to the critical current of the form $\delta I_c\sim
(\Delta^2/E_{\rm Th}) \ln(\Delta/E_{\rm Th})$, nonanalytic in the
limit of the short normal region.

Summarizing the existing knowledge, one may write  for the critical
current of an SNS structure
\begin{equation}\label{ScalingRelation}
I_{c}=\frac{g\Delta}{e}\, \mathcal{I}(\kappa),\quad\,\,\,\,
\kappa\equiv\frac{\Delta}{E_{\mathrm{Th}}}\,,
\end{equation}
where $g$ is the conductance of normal region and
$\mathcal{I}(\kappa)$ is a universal scaling function of the
junction's dimensionless length $\kappa=(L/\xi)^2$. In the limit of
long junction, $\kappa \gg 1$, this function is given by
\cite{Wilhelm} $\mathcal{I}(\kappa)\approx 10.82\ \kappa^{-1}$. On
the other hand, for shorter junctions, $\kappa \ll 1$, it is
\begin{equation}\label{ScalingFunction}
\mathcal{I}(\kappa)= \mathcal{I}(0) -
a\kappa\ln(b/\kappa)+O(\kappa^2) \, ,
\end{equation}
where $\mathcal{I}(0)=2.08$, see, e. g.,
Ref.~\onlinecite{Beenakker}, and $a=0.31$, while $b=2.84$. The first
term on the right-hand side (r.h.s.) represents the current carried
by the Andreev states and was discussed by Kulik and Omelyanchuk
\cite{KO} and Beenakker \cite{Beenakker}. The second nonanalytic
term originates from the above--gap extended states.  This term is
the main result of the present paper. In the rest of the paper, we
derive it and compare our analytical result with the
existing~\cite{Wilhelm} numerical data.

To approach the problem analytically, we employ the Usadel equation
\cite{Usadel} for the matrix Green function $\hat G(\varepsilon,x)$
of the quasi--1D disordered normal region,
\begin{equation}
                                                    \label{Usadel}
D\partial_x(\hat{G}\partial_x\hat{G})+
\mathrm{i}\varepsilon [\hat{\sigma}_{3},\hat{G}]=0\, ,
\end{equation}
where the Green function is subject to the constraint $\hat G^2=\hat
1$. At the two NS interfaces, the Green functions $\hat
G(\varepsilon,\pm L/2)$ are given by those of the BCS
superconductors with the order parameter $\Delta$, maintained at the
phase difference $\phi$. The suppression of the order parameter
within the bulk superconducting leads may be safely disregarded as
long as the transverse directions of the normal region are less than
the coherence length, $\xi$. The same assumption justifies keeping
only the lowest transverse mode  and thus the use of the 1D form of
the Usadel equation. This should be contrasted to the case of the
long junction $L\gg\xi$, where fully self--consistent calculation is
required and the approximation with the constant order parameter
within the normal region is not appropriate.

The Green function may be parametrized by the two complex angles
$\theta(\varepsilon,x)$ and $\varphi(\varepsilon,x)$ as
\begin{equation}\label{thetaphi}
\hat G =\sin\theta\cos\varphi\, \hat\sigma_1 +
\sin\theta\sin\varphi\, \hat\sigma_2 +\cos\theta\,\hat\sigma_3\, .
\end{equation}
Employing this parametrization and introducing the rescaled
coordinate $x/L\to x$, we rewrite the Usadel equation (\ref{Usadel})
as
\begin{equation}\label{Usadel_NormalWire}
\partial^2_x \theta + \omega^{2}\sin\theta = J^2
\cos\theta\left(\sin \theta\right)^{-3} \,,
\end{equation}
where $\omega^2\equiv 2\,\mathrm{i}\varepsilon/E_{\rm Th}$, and the
spectral current density $J(\varepsilon)$ is given by
\begin{equation}
                \label{J}
J=\sin^2\theta\, \partial_x\varphi\, .
\end{equation}
One needs to solve Eq.~(\ref{Usadel_NormalWire}) with the boundary
conditions
\begin{equation}\label{BoundaryConditions}
\tan\theta(\varepsilon,\pm
1/2)=\mathrm{i}\Delta/\varepsilon.
\end{equation}
The spectral current density may then be determined from the
condition of having the fixed phase difference:
$\varphi(\varepsilon,1/2)-\varphi(\varepsilon,-1/2)=\phi$. Finally,
the   Josephson current  is found as
\begin{equation}\label{CFR_Definition}
I(\phi)=\frac{g}{e}
\int\limits_{0}^{\infty}\!\mathrm{d}\varepsilon\,
\tanh\left(\frac{\varepsilon}{2T}\right)\ \mathrm{Im} \,
J(\varepsilon,\phi)\, .
\end{equation}
For the sake of illustration, we shall first execute this program in
the short junction limit, $L\sim\omega \to 0$. Equation
(\ref{Usadel_NormalWire}) with $\omega=0$ may be easily integrated,
resulting in
\begin{equation}
                                            \label{Usadel_Solution}
\cos\theta(\varepsilon,x)=\cos\theta_{0}\cos\left(\frac{Jx}{\sin\theta_{0}}\right)\,
,
\end{equation}
where the integration constant is
$\theta_{0}(\varepsilon)=\theta(\varepsilon,0)$. Integrating then
Eq.~(\ref{J}) and employing the boundary condition for the phase
$\varphi$, one obtains
\begin{equation}\label{RelationTwo}
\tan(\phi/2)=\frac{1}{\sin\theta_{0}}\,
\tan\left(\frac{J}{2\sin\theta_{0}}\right)\, .
\end{equation}
This last equation along with Eq.~(\ref{Usadel_Solution}) taken at
the NS interface, $x=1/2$, constitute the system of the two {\em
algebraic} equations for the two unknown quantities:
$J(\varepsilon)$ and $\theta_0(\varepsilon)$. Such an algebraic
problem may be easily solved, resulting in the following expression
for the imaginary part of the spectral current:
\begin{equation}
                                          \label{J_ImaginaryPart}
\mathrm{Im}\,J = \frac{\pi\Delta
\cos(\phi/2)}{\sqrt{\varepsilon^2-\Delta^2\cos^2(\phi/2)}}
\end{equation}
for $\Delta\cos(\phi/2)<\varepsilon<\Delta$, and $\mathrm{Im}\,J = 0$
otherwise. The fact that $\mathrm{Im}\,J$ vanishes for $\varepsilon
>\Delta$ is an artefact of the approximation $E_{\rm Th}\to \infty$,
already mentioned in the beginning of this paper. Equation
(\ref{J_ImaginaryPart}) is in perfect agreement with the result of
Beenakker \cite{Beenakker} based on the consideration of the Andreev
bound states in the diffusive normal region. One concludes that the
Josephson current in the limit $L\to 0$ is entirely given by the
Andreev states residing inside the superconducting gap. Employing
Eqs.~(\ref{CFR_Definition}) and (\ref{J_ImaginaryPart}) one arrives
at the well--known result \cite{KO,Beenakker} for the
zero--temperature Josephson current  of the short SNS junction,
\begin{equation}\label{CFR-KO}
I(\phi)=I_{o}\, \cos(\phi/2)\,\mathrm{arctanh}
\big[ \sin(\phi/2)\big]\, .
\end{equation}
with $I_{o}=\pi g \Delta/e$. At $\phi_{\rm max}= 1.97$, this function reaches its maximum value
$I_c\equiv I(\phi_{\rm max})=2.08g\Delta/e$, which defines the
coefficient $\mathcal{I}(0)=2.08$ in Eq.~(\ref{ScalingFunction}).

Having established the limit of an extremely short junction, we turn
now to our main subject: the finite--length correction to the
critical current. As was mentioned in the introductory section and
will be proven below, the largest correction originates from the
parametrically wide range of energies well above the superconducting
gap: $\Delta<\varepsilon<E_{\rm Th}$. One may notice that in this
energy interval, the solution for $\theta(\varepsilon)$ must be of
the order of $\Delta/\varepsilon \ll 1$, allowing for
small--$\theta$ expansion in Eq.~(\ref{Usadel_NormalWire}),
\begin{equation}
                                      \label{Usadel_Linearized1}
\partial^2_x\theta+ \omega^{2}\theta = J^{2} \theta^{-3}\, .
\end{equation}
Within the same approximation, the boundary conditions
(\ref{BoundaryConditions}) read
$\theta(\varepsilon,\pm1/2)=\mathrm{i}\Delta/\varepsilon$, while
Eq.~(\ref{J}) for the spectral current density takes the form
$J=\theta^2\partial_x\varphi$.

Equation \eqref{Usadel_Linearized1} may be solved exactly, leading
to
\begin{equation}
                                             \label{TetaSolution}
\theta^{2}(\varepsilon,x)=\theta^{2}_{0}\cos^{2}(\omega x)
+\frac{J^{2}}{\theta^{2}_{0}}\, \frac{\sin^{2}(\omega
x)}{\omega^{2}}\, ,
\end{equation}
where $\theta_{0}(\varepsilon)=\theta(\varepsilon,0)$ is the
integration constant, similar to the one in
Eq.~\eqref{Usadel_Solution}. Substituting the solution
Eq.~(\ref{TetaSolution}) into the equation for the spectral current
and integrating over the coordinate, one finds
\begin{equation}\label{RelationFour}
\tan(\phi/2)=\frac{J}{\theta^{2}_{0}}\,
\frac{\tan\left(\omega/2\right)}{\omega}\, .
\end{equation}
Notice  that in the limit $\omega\rightarrow 0$
Eqs.~\eqref{TetaSolution} and \eqref{RelationFour} reduce  to the
small $\theta_0$ and small $J$ limit of
Eqs.~\eqref{Usadel_Solution} and \eqref{RelationTwo},
respectively, as they should.

Taking Eq.~\eqref{TetaSolution} at the NS interface, $x=1/2$, and
employing the boundary conditions, one obtains the algebraic
relation between yet unknown quantities $\theta_0$ and $J$. The
second relation between these two quantities is provided by
Eq.~(\ref{RelationFour}). Resolving these two algebraic relations,
one finds for the spectral current density
\begin{equation}
                                      \label{SpectralCurrent2}
J=-\frac{\Delta^2}{\varepsilon^2} \,
\frac{\omega}{\sin\omega}\,\sin\phi\,\, .
\end{equation}
In the short junction limit, $\omega\to 0$, this expression is real,
which is in agreement with our previous finding that $\mathrm{Im}\,
J = 0$ for $\varepsilon > \Delta$. However, for a finite--length
junction (recalling that $\omega=\sqrt{2\mathrm{i}\varepsilon/E_{\rm
Th}}$) one finds the non-zero imaginary part of the spectral
current: $\mathrm{Im}\,J\approx-\Delta^2/(3 E_{\rm Th} \varepsilon)$
in the energy range $\Delta<\varepsilon< E_{\rm Th}$. For larger
energies, $\varepsilon>E_{\rm Th}$, Eq.~\eqref{SpectralCurrent2}
predicts exponentially decaying
$\mathrm{Im}\,J\propto\exp(-\sqrt{\varepsilon/E_{\mathrm{Th}}})$.
Indeed, as was discussed in the introductory part, the energies
above the Thouless energy should not contribute to the supercurrent.
Note that the tail of the spectral current found in the
Eq.~(\ref{SpectralCurrent2}) at the energies above the gap is in
perfect agreement with the results of the numercial
studies~\cite{Yeyati}. Integrating Eq.~(\ref{SpectralCurrent2}) over
the energy according to Eq.~(\ref{CFR_Definition}), one finds the
finite--size correction to the $T=0$ Josephson current,~\cite{foot}
\begin{equation}\label{Is-Correction-Final}
\delta I(\phi) =
-I_{o}\left(\frac{\kappa}{3\pi}\right)\ln(b/\kappa)\, \sin\phi \, .
\end{equation}
This correction modifies the value of the critical current, $\delta
I_c=\delta I(\phi_{\rm max})$, with $\phi_{\rm max}$ quoted after
Eq.~(\ref{CFR-KO}). This way we obtain the second term in the r.h.s.
of Eq.\eqref{ScalingFunction} with the coefficient $a=(\sin\phi_{\rm
  max})/3=0.31$.

\begin{figure}
\includegraphics[width=8.5cm]{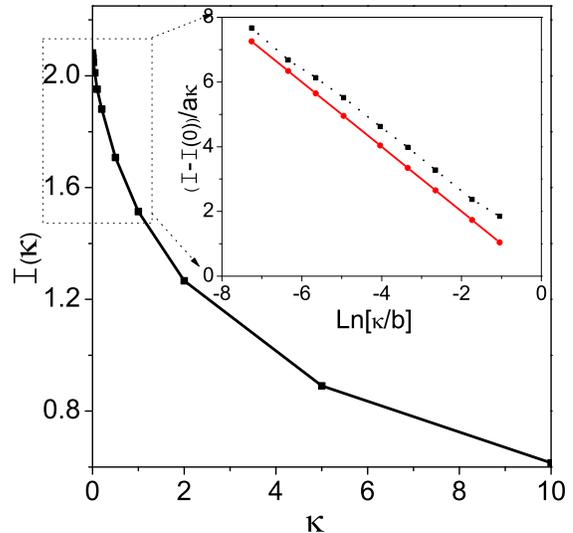}
\caption{(Color online) Numerical results of Ref.~\cite{Wilhelm} for the scaling
  function $\mathcal{I}(\kappa)$. Inset: to facilitate the comparison
  of numerical results (Ref.~\onlinecite{Wilhelm}) (data points connected by a
  dashed line) with analytical theory [solid straight line corresponding
  to Eq.~(\ref{ScalingFunction})], the data are replotted in
  co-ordinates $[\mathcal{I} - \mathcal{I}(0)]/a\kappa$ vs
  $\ln(\kappa/b)$ for the region $\kappa\leq 1$.
  \label{Fig:Crossover}}
\end{figure}

We comment here on the additional approximation made while deriving
Eq.~(\ref{Is-Correction-Final}). Our solution strategy was based on
the perturbation theory for the Usadel equation over $L/\xi\ll1$
exploring smallness on the phase $\theta(x,\varepsilon)$ in the
energy range $\Delta<\varepsilon<E_{\mathrm{Th}}$. But we have used
rigid boundary conditions (\ref{BoundaryConditions}) ignoring
corrections to them coming from the finiteness of the normal region
link. Although these corrections exist, they do not bring any new
contributions \textit{non-analytical} in $L/\xi$ to
Eq.~(\ref{Is-Correction-Final}).

The crossover between the limits of short and long junctions was
recently studied numerically by Dubos {\sl et al.} \cite{Wilhelm}.
In Fig.~\ref{Fig:Crossover}, we use the data of
Ref.~\onlinecite{Wilhelm} to plot the scaling function
$\mathcal{I}(\kappa)$. At small $\kappa$, we expect the linear
dependence of $[\mathcal{I} - \mathcal{I}(0)]/\kappa$ on $\ln
\kappa$, see Eq.~(\ref{ScalingFunction}). The numerical data
\cite{Wilhelm} indeed agree with this expectation, as shown in the
inset to Fig.~\ref{Fig:Crossover}.

All the above consideration was based on the assumption that the
interfaces between normal metal and superconductor are perfectly
transparent. In most experimental configurations, SNS junctions
always contain the potential barriers, an insulating layer I, at the
SN interfaces. Thus it is of interest to investigate the influence
of the extended quasiparticle energy levels on the Josephson current
in the realistic SINIS configuration. In what follows, we will
constraint ourselves to the case of symmetric SINIS junction with
tunnel barrier conductance $g_{T}$ being much smaller then $g$.

The physics of the Josephson effect in the SINIS junction is
controlled by the dimensionless interface parameter
$\gamma_{\mathrm{eff}}=(L/\xi)^{2}(g/g_{T})$. In the so-called
incoherent regime $\gamma_{\mathrm{eff}}\gg1$, the SINIS junction
may be viewed as the two tunnel junctions connected in series with
the current-phase relation in the form
$I(\phi)=I_{c}\sin(\phi/2)$~\cite{Review}. The most interesting case
is the coherent regime $\gamma_{\mathrm{eff}}\ll1$ which we consider
below.

To calculate the supercurrent for the SINIS junction, we start from
the point contact limit and follow the same steps
(\ref{Usadel_NormalWire})-(\ref{RelationTwo}) as for SNS case but
with one significant difference, namely the boundary conditions for
Usadel equation. Instead of rigid boundary condition
(\ref{BoundaryConditions}), we apply conditions appropriate for this
case,\cite{KupriyanovLukichev}
\begin{subequations}\label{KL-Boundary}
\begin{equation}
J=\frac{2g_{T}}{g}\sin\theta_{B}\sin\theta_{S}\sin\frac{\phi-\varphi_{B}}{2},
\end{equation}
\begin{equation}
\!\!\partial_{x}\theta|_{x=\frac{1}{2}}=\frac{2g_{T}}{g}
\left[\sin\theta_{S}\cos\theta_{B}\cos\frac{\phi-\varphi_{B}}{2}-
\sin\theta_{B}\cos\theta_{S}\right],
\end{equation}
\end{subequations}
where $\theta_{B}(\varepsilon)=\theta(\varepsilon,\pm1/2)$,
$\varphi_{B}=\varphi(\varepsilon,1/2)$ and
$\cos\theta_{S}=\varepsilon/\sqrt{\varepsilon^{2}-\Delta^{2}}$. As
long as the tunnel conductance is small $g_{T}\ll g$ the phase
function $\varphi(\varepsilon,x)$ changes abruptly at the tunnel
barriers from its values $\pm\phi/2$ inside the superconductors to
$\pm\varphi_{B}$ at the boundaries of the junction and remains small
inside the normal region $|\varphi(\varepsilon,x)|\ll1$. These
observations implies that in the boundary conditions
\eqref{KL-Boundary} one can safely put $\varphi_{B}=0$. As follows
from the Eq.\eqref{KL-Boundary} the spectral current $J$ is small as
$\sim g_{T}/g\ll1$. Therefore, according to the solution
(\ref{Usadel_Solution}) of the Usadel equation the
$\theta(\varepsilon,x)$ is essentially coordinate independent
$\partial_{x}\theta(x,\varepsilon)=0$, provided
$E_{\mathrm{Th}}\gg\Delta g/g_{T}$. In this case the boundary
conditions \eqref{KL-Boundary} represent the closed system of
algebraic equations for unknown functions $\theta_{B}$ and $J$,
which can be solved in terms of $\phi$ and $\theta_{S}$. As a result
the imaginary part of the spectral current takes the form
\begin{equation}\label{SpectralCurrent3}
\mathrm{Im}J=\frac{g_{T}}{g}
\frac{\Delta^{2}\sin\phi}{\sqrt{\Delta^{2}-\varepsilon^{2}}
\sqrt{\varepsilon^{2}-\Delta^{2}\cos^{2}(\phi/2)}}.
\end{equation}
Combining this with Eq.(\ref{CFR_Definition}) at zero temperature,
we find the Josephson current-phase relation in the
form~\cite{KupriyanovLukichev}-\cite{BrinkmanGolubov}
\begin{equation}
I(\phi)=I_{o}(2g_{T}/\pi g)K[\sin(\phi/2)]\sin\phi,
\end{equation}
where $K(x)$ is the complete elliptic integral of the first kind.
Let us compare this result with Eq.(\ref{CFR-KO}). First of all, we
observe that the presence of the tunnel barriers changes the
current--phase relation, but preserves its essential properties, for
example nonanalyticity at phase $\phi=\pi$. Secondly, the amplitude
of the critical current is suppressed by the small parameter
$g_{T}/g\ll1$ compared to its value, Eq.(\ref{CFR-KO}). Similar
changes occur with the Josephson current correction $\delta I$: we
get modified phase-$\phi$ dependence, the amplitude of the current
correction is suppressed due to the tunnel barriers, but the
correction itself remains nonanalytical in $L/\xi$. The estimate
shows that the amplitude of the Josephson current correction in the
symmetric SINIS junction is
\begin{equation}
\delta I_{c}\propto-I_{o}\left(g_{T}/g\right)
\kappa\ln(1/\kappa).
\end{equation}

Let us discuss now some characteristic features of the above--gap
contribution to the Josephson current found in this paper: (i) The
power--law tail of the spectral current density, $\mathrm{Im}\,J\sim
1/\varepsilon$, results in the nonanalytic length dependence of the
critical current, $\delta I_c\sim L^2\ln L$. (ii) The supercurrent
carried by the high--energy extended states is negative, i.e., it
flows in the direction opposite to the current produced by the
Andreev bound states. (iii) Unlike the dependence on length, the
phase dependence of the found contribution to Josephson current is
not singular, $\delta I(\phi) \sim\sin\phi$. This should be
contrasted with the contribution coming from the Andreev bound
states, Eq.~(\ref{CFR-KO}). The latter contains (at any $L$)  a
nonanalytic phase dependence, $I(\phi)\sim(\pi-\phi)\ln|\pi-\phi|$,
at $\phi\approx\pi$. The origin of such a difference between the two
contributions is in the evolution of the Andreev states with the
phase $\phi\,$: unlike above--gap states, the lowest of the bound
states depends on the phase in a peculiar way, ``touching'' the
Fermi level, $\varepsilon=0$, at $\phi=\pi$. (iv) The occupation of
the states contributing to the $L^2\ln L$ dependence of the
Josephson current hardly changes with temperature, as long as
$T\lesssim\Delta(T)$. As a result, the temperature dependence of the
second term in the r.h.s. of Eq.~\eqref{ScalingFunction} comes
solely from the temperature--dependent superconducting gap
$\Delta(T)$.  (v) At a temperature close to the critical one: $T_c -
T\ll T_c< E_{\rm Th}$, the short SNS bridge is in the regime where
$\Delta(T)\ll T$. Performing energy integration according to
Eq.~(\ref{CFR_Definition}), one finds for the Josephson current:
\begin{equation}\label{Is-highT}
I(\phi)+\delta I(\phi) = \frac{g\Delta^{2}(T)}{e} \left(\frac{\pi}{4T}-
\frac{\ln (E_{\rm Th}/T_c)}{3E_{\rm Th}}
\right) \sin\phi \,
,
\end{equation}
where the first term in brackets originates from the Andreev states,
while the second one originates from the high energy extended
states. The nonanalytic length dependence exists in this case as
well.

Although our calculations were carried out for the case of the
diffusive normal region, it is clear that the effect is rather
generic. In particular, with the proper redefinition of the Thouless
energy (and possibly with different numerical constants),
Eqs.~\eqref{ScalingRelation} and \eqref{ScalingFunction} should hold
for ballistic SNS structures as well~\cite{Sharov}. Moreover,
nonanalytical correction survives even in the case of SINIS junction
but with suppressed amplitude due to small tunnel barriers
conductance. Also the critical current of superconducting weak links
of the type SS'S should contain the nonanalytic term, originating
from the high--energy states, and numerical studies \cite{Yeyati}
only support this conclusion.

We are grateful to H. Courtois, F.K. Wilhelm and Levy Yeyati for useful
discussions, and H. Courtois for providing us with the results of their
numerical calculation \cite{Wilhelm}, which enabled us to perform
the comparison presented in Fig.~\ref{Fig:Crossover}. This work
is supported by NSF grants DMR 02-37296, DMR 04-39026 DMR 0405212,
and EIA 02-10736.  A.~K. is also supported by the A.P. Sloan
foundation.


\begin{thebibliography}{99}

\bibitem{Nano1} Y. Doh, J.A. Dam, A.L. Roest, E. Bakkers, L.P.
Kouwenhoven, S. Franceschi, Science \textbf{309}, 272, (2005).

\bibitem{Nano2} A. Kasumov, M. Kociak, M. Ferrier, R. Deblock, S.
Gueron, B. Reulet, I. Khodos, O. Stephan, H. Bouchiat, Phys. Rev.
B \textbf{68}, 214521, (2003).

\bibitem{Nano3} B.J. van Wees, H. Takayanagi, \textit{Proceeding of the
NATO Advanced Study Institute on Mesoscopic Electron Transport,
edited by L.P.Kouwenhoven }, p.469, (1996).

\bibitem{Nano4} B.J. Wees, Physics World, 9(11), 41, (1996).

\bibitem{Review} A.A. Golubev, M.Yu. Kupriyanov, E.Il'ichev,
Rev. Mod. Phys. \textbf{76}, 411, (2004).

\bibitem{Beenakker} C.W.J. Beenakker, Phys. Rev. Lett.
\textbf{67}, 3836, (1991) and \textit{Transport Phenomena in
Mesoscopic Systems}, edited by H. Fukuyama and T. Ando (Springer,
Berlin, 1992).

\bibitem{Heikkila} T.T. Heikkila, J. Sarkka and F.K. Wilhelm, Phys.
Rev. B \textbf{66}, 184513, (2002).

\bibitem{KO} I.O. Kulik, A.N. Omelyanchuk, Sov. J. Low. Temp.
Phys, \textbf{3}, 459, (1977) and \textbf{4}, 142, (1978).

\bibitem{Schon} F.K. Wilhelm, A.D. Zaikin, G.Schon, J. Low Temp.
Phys. \textbf{106}, 305, (1997).

\bibitem{Zaikin} W. Belzig, F.K. Wilhelm, C. Bruder, G. Sh\"{o}n,
A.D. Zaikin, Superlattices Microstruct. \textbf{25}, 1251, (1999).

\bibitem{Wilhelm} P. Dubos, H. Courtois, B. Pannetier, F.K. Wilhelm,
  A.D. Zaikin and G. Sch\"{o}n, Phys. Rev. B \textbf{63}, 064502,
  (2001).

\bibitem{Usadel} K.D.
Usadel, Phys. Rev. Lett, \textbf{25}, (1970).

\bibitem{Andreev} A.F. Andreev, Sov. Phys. JETP, \textbf{19}, 1228,
(1964).

\bibitem{foot} Finding the coefficient $b=2.84$ inside the logarithm
  requires more careful consideration of the energy range
  $\varepsilon\approx \Delta$, where $\theta$ is not small and one has
  to work with Eq.~(\ref{Usadel_NormalWire}), rather than
  Eq.~(\ref{Usadel_Linearized1}); A. Levchenko, A. Kamenev, L.
  Glazman, unpublished.

\bibitem{KupriyanovLukichev} M.Yu. Kupriyanov, V.F. Lukichev, Sov. Phys. JETP, \textbf{67},
1163, (1988)

\bibitem{BrinkmanGolubov} A. Brinkman, A.A. Golubov,  Phys. Rev. B
\textbf{61}, 11297, (2000)

\bibitem{Yeyati} A. Levy Yeyati, A. Martín-Rodero, and F. J. García-Vidal,
Phys. Rev. B, \textbf{51}, 3743 (1995)

\bibitem{Sharov} S.V. Sharov, A.D. Zaikin, Phys. Rev. B
\textbf{71}, 014518, (2005)
\end{thebibliography}
\end{document}